\documentclass[letterpaper,11pt]{article}
\pdfoutput=1 

\usepackage{jheppub} 

\usepackage{tikz}
\usepackage{dsfont}
\usepackage{amsmath}
\def\ba{\begin{align}}\def\ea{\end{align}}
\def\beq{\begin{eqnarray}}\def\eeq{\end{eqnarray}}
\def\be{\begin{equation}}\def\ee{\end{equation}}
\def\ben{\begin{equation}}
\def\een{\end{equation}}
\def\bea{\begin{eqnarray}}
\def\eea{\end{eqnarray}}

\def\vx{{\vec{x}}}
\def\vm{{\vec{m}}}
\def\vk{{\vec{k}}}
\def\vy{{\vec{y}}}
\def\vomega{{\vec{\omega}}}

\def\cO{{\cal{O}}}
\def\tcO{\tilde{\cal{O}}}

\def\fr{\frac}
\def\lan{\langle}
\def\ran{\rangle}

\def\eq#1{(\ref{#1})}

\def\G{\Gamma}
\def\A{\alpha}
\def\B{\beta}
\def\bz{{\bar z}}
\def\bw{{\bar w}}

\title{Operators in the Internal Space and Locality}

\author{Hardik Bohra $^1$,}
\author{Sumit R. Das$^1$,}
\author{Gautam Mandal$^2$,}
\author{Kanhu Kishore Nanda $^2$,}
\author{Mohamed Hany Radwan$^1$,}
\author{Sandip P. Trivedi$^2$}

\affiliation{$^1$Department of Physics and Astronomy, University of Kentucky, Lexington, KY 40506, U.S.A.}
\affiliation{$^2$Department of Theoretical Physics, Tata Institute of Fundamental Research, Mumbai 400005, INDIA}

\emailAdd{bohra.hardik@uky.edu}
\emailAdd{das@pa.uky.edu}\emailAdd{mandal@theory.tifr.res.in}
\emailAdd{kanhu.nanda@tifr.res.in}
\emailAdd{m.radwan@uky.edu}
\emailAdd{sandip@theory.tifr.res.in}

\abstract{Realizations of the holographic correspondence in String/M theory typically involve spacetimes of the form $AdS \times Y$ where $Y$ is some internal space which geometrizes an internal symmetry of the dual field theory, hereafter referred to as an ``$R$ symmetry". It has been speculated that areas of Ryu-Takayanagi surfaces anchored on the boundary of a subregion of $Y$, and smeared over the base space of the dual field theory, quantify entanglement of internal degrees of freedom. A natural candidate for the corresponding operators are linear combinations of operators with definite $R$ charge with coefficients given by the ``spherical harmonics'' of the internal space: this is natural when the product spaces appear as IR geometries of higher dimensional AdS spaces. We study clustering properties of such operators both for pure $AdS \times Y$ and for flow geometries, where $AdS \times Y$ arises in the IR from a different spacetime in the UV, for example higher dimensional AdS or asymptotically flat spacetime. We show, in complete generality, that the two point functions of such operators separated along the internal space obey clustering properties at scales larger than the $AdS$ scale. For non-compact $Y$, this provides a notion of approximate locality. When $Y$ is compact, clustering happens only when the size of $Y$ is parametrically larger than the $AdS$ scale. 
This latter situation is realized in flow geometries where the product spaces arise in the IR from an asymptotically AdS geometry at UV, but not typically when they arise near black hole horizons in asymptotically flat spacetimes. We discuss the significance of this result for entanglement and comment on the role of color degrees of freedom. }

\begin{document}

\begin{flushright}
\end{flushright}

\maketitle
\flushbottom

\section{Introduction}
\label{sec:intro}


The past several decades have uncovered a deep relationship between entanglement in holographic quantum field theories and emergence of smooth gravitational bulk \cite{review}. This is best understood in AdS/CFT realizations of holography. The kind of entanglement which is most widely discussed in this context is entanglement in base space, i.e. entanglement of a sub-region of space $A$ on which the field theory is defined, with its complement. In this case, generalizations of the Ryu-Takayangi proposal \cite{rt} -\cite{ew} lead to the definition an entanglement wedge in the bulk: 
the domain of dependence of a  bulk hypersurface bounded by the generalised RT surface corresponding to $A$.  Entanglement wedge reconstruction then reconstructs operators in the entanglement wedge from field theory operators in $A$
\cite{wedge,wedge2}. It has been in fact recently emphasized that $A$ need not be a part of the boundary, and could be some part of the bulk itself - and these more general situations possibly realize entanglement wedge reconstruction in spaces which are not asymptotically $AdS$ \cite{bousso} - in fact the proof of the RT formula in \cite{lm} already deals with such regions. 

In microscopic realizations of $AdS/CFT$, the $AdS$ space is usually accompanied with an internal space, e.g. in $AdS_5 \times S^5$, $AdS_4 \times S^7$ or $AdS_3 \times S^3 \times T^4$. The internal space geometrizes internal global symmetries of the field theory which we call R symmetry. From the point of view of the bulk space-time, the internal space (e.g. $S^5$) is on par with  $AdS_5$. It is therefore natural to ask if there is an entanglement wedge reconstruction which reconstructs a region of the bulk from a sub-region of the internal space. From the point of view of the dual field theory this kind of entanglement cannot be the usual base space entanglement. 

A first step in understanding this issue would be to understand if there is a meaning of the area of a Ryu-Takayanagi surface which is anchored on the boundary of a sub-region of the internal space at the AdS cutoff boundary, and smeared along the $AdS$ spatial directions. This question was first investigated in \cite{shiba}. An interpretation of such a RT surface is somewhat confusing in view of a result due to Graham and Karch \cite{gk} which states that any such codimension two RT surface which goes all the way to the boundary of AdS has to end on a codimension three surface which is itself an extremal surface in the internal space, e.g. if the internal space is a sphere, the anchor of such a surface is the equator. In a cutoff $AdS$ space such a surface {\em can} end on an arbitrary subregion, and one may try to find a meaning of such surfaces. A cutoff $AdS$ space corresponds to a boundary field theory with a finite UV cutoff. \cite{shiba} considered RT surfaces in $AdS_5 \times S^5$ with a cutoff $AdS$, ending on a cap with some lattitude angle $\theta_0$. Their (numerical)  solutions showed that for small $\theta_0$ the area of the RT surface is proportional to the {\em volume} of the cap. \cite{shiba} proposed that this quantity measures the entanglement between a $SU(M)$ and $SU(N-M)$ subsectors of the dual $SU(N)$ gauge theory, with $M$ being proportional to the volume of the cap. In general, it is not clear how to make a gauge-invariant sense of such a division of the degrees of freedom, though this can be made a lot more precise in the Coulomb branch \cite{shiba,ku}. 

A somewhat different proposal was made by \cite{ku}. Let us start with the primary operators in the Yang-Mills theory. They have definite R-symmetry quantum numbers. In the $S^5$ example, these operators $\cO_{l,\vm} (x^\mu)$ are labelled by the $SO(6)$ quantum numbers $(l,m_1 \cdots m_4)$. Here $l$ determines the value of the Casimir and $m_i$ are the "magnetic" quantum numbers, while $x^\mu$ denote the base space coordinates. The idea is to then take linear combinations of these operators with the spherical harmonics $Y_{l,\vm} (\theta_i)$ as coefficients to form operators which are now functions of the base space  coordinates $x^\mu$ and labelled by the angles on the sphere $\theta_i$,
\ben
\cO (x^\mu,\theta_i) \equiv \sum_{l,\vm} Y_{l,\vm}(\theta_i) \cO_{l,\vm} (x^\mu)
\label{0-1}
\een
The proposal is that the RT surface which ends on a subregion $A$ of the sphere $S^5$ at the cutoff boundary measures the von Neumann entropy associated with the reduced density matrix which evaluates expectation values of operators belonging to a subalgebra constructed by considering
$\cO (x^\mu,\theta_i)$ where are $\theta_i$ are now restricted to lie in this region $A$, and taking products and sums of operators of this type. This kind of construction appears in discussions of entanglement entropies in non-commutative field theories \cite{noncommutative} where e.g. the sum in (\ref{0-1}) extends to some finite value of $l = l_{max}$ to construct a function on a fuzzy sphere. Such operators have also been considered as interpretations of areas of extremal surfaces in D0 brane backgrounds \cite{anous}.  

Unlike the proposal of \cite{shiba}, the projections used in definining this reduced density matrix do not {\em explicitly} involve the color space. However, as we will see below, the color degrees of freedom are indirectly involved: restriction of operators like (\ref{0-1}) must come from a restriction of some subset of the color degrees of freedom, albeit in a gauge invariant fashion.

In a recent paper \cite{dkkmrt}, some of us looked at this issue from a different viewpoint. We looked at geometries where such product space-times appear as IR geometries of a higher dimensional asymptotically $AdS$ space-time. Well known examples include extremal charged black holes (branes) in $AdS_{d+1}$ whose near-horizon geometry is $AdS_2 \times S^{d-1}$ ($AdS_2 \times R^{d-1}$), magnetic branes which flow from $AdS_5$ to $AdS_3 \times R^2$ \cite{kraus,almuhairi,jerome} \footnote{In the absence of supersymmetry these solutions are generically unstable \cite{jerome}}, as well geometries with "boomerang" RG flows where such a product space appears at an intermediate scale \cite{jerome2}. Thus some of the base space directions of the field theory in the UV become internal dimensions of a putative field theory dual to the IR geometry.

We considered RT surfaces which completely cover the spatial directions which will become a part of the lower dimensional $AdS$, but cover a subregion $A$ of the remaining directions which appear as internal directions in the IR. Since the geometry is asymptotically $AdS$, we know the meaning of the area of such a surface. In terms of the UV field theory, this is the entropy associated with the subalgebra of local operators restricted to $A$. It was found that when the region $A$ is large compared to the scale of the RG flow (which is the horizon radius for the extremal black hole or the magnetic field for magnetic brane) the corresponding RT surface proceeds almost radially inward, without traversing much of the directions along $A$, till they reach the IR region. This raises the possibility that the leading size dependence of the entropy comes from the IR part of the geometry and has an interpretation in a theory which is dual to the IR geometry. Such a putative dual must reside on a holographic screen which is located at the junction of the UV and IR geometries. The specific examples studied in \cite{dkkmrt} lead to the following general picture. When the IR geometry is $AdS_2 \times S^{d-1}$ or $AdS_2 \times R^{d-1}$, as in extremal Reissner-Nordstrom black hole or black brane, the warp factor plays a crucial role, and one can extract the IR part of the entropy in a way which is completely insensitive to the location of the holographic screen on which the IR field theory lives. The result is proportional to the volume of the subregion $R$, in units with the scale of the RG flow, which is the horizon radius $r_h$. 

The situation is different for cases where the IR geometry contains a $AdS_{d+1-n}$ with 
$(d-n) > 1$. In these latter cases the location of the holographic screen becomes relevant. Consider e.g. magnetic branes dual to $N=4$ Yang-Mills in $3+1$ dimensional flat space in the presence of a constant magnetic field $B = F_{12}$. In this case, the RT surface reaches the Poincare horizon when the width of the strip, $l$,  reaches a finite value, and this scale is also the scale at which the geometry can be approximated by a product space. When the width exceeds this value, the RT surfaces go straight into the horizon and their areas stop depending on the width. Indeed, an analysis of RT surfaces in the product geometry $AdS_{d+1-n}\times R^n$ which end on the boundary of a strip with width $l$ extending along one of the $R^n$ directions shows this behavior as well for $(d-n) > 1$. Such a surface reaches the horizon when $l \sim 1/{(d-n-1)}$. The area of the surface is now proportional to $r_{UV}^{d-n-1}$, the cutoff of the $AdS_{d+1-n}$. When embedded in  the higher dimensional $AdS$ we must have $r_{UV} \sim l_{RG}$ (e.g. $r_{UV} \sim 1/B$ for magnetic branes).

The case $d-n=1$ is special since this is the only situation where the geometry has a horizon with non-vanishing area. Accordingly, the dual state has a non-vanishing classical entropy. The extensive behavior is in fact this classical entropy. For $d-n > 1$ the classical area of the horizon vanishes, so that the dual state is a pure state.

These results are consistent with the proposal in \cite{ku}. Consider for example a geometry which flows from Poincare patch $AdS_{d+1}$ in the UV to $AdS_{d+1-n} \times R^n$ in the IR. The radial coordinate is denoted by $r$, while the other coordinates are denoted by $(x^\mu, y^i)$ with $\mu = 0,\cdots (d-n)$ and $i = 1 \cdots n$. In the IR, the latter directions, $y^i$ split off into a space $R^n$ \footnote{Similarly for geometries which describe a flow from global $AdS_{d+1}$ to global $AdS_{d+1-n} \times S^n$, $x^\mu$ denote the time and $(d-n)$ angles $\phi_a$ while $y^i$ denote $n$ angles on a $S^n$ which splits off in the IR. }. In the dual field theory this is a rather unconventional RG flow from a theory with $(d+1)$ dimensional base space-time to another theory with a $(d+1-n)$ dimensional base space-time whose coordinates will be denoted by $x^\mu$. The primary operators of the UV theory are of the form $\cO(x^\mu,y^i)$ which may be decomposed in terms of the momenta $k_i$ along the $y^i$ directions, $\cO_{k_i} (x^\mu)$. In the IR of the bulk, $k_i$ are the Kaluza-Klein momenta associated with the internal space. In a putative dual theory, the primary operators are indeed $\cO_{k_i} (x^\mu)$ with dimension which depends on $k_i$. For global $AdS_{d+1-n} \times S^n$ the role of $k_i$ are played by the angular momenta quantum numbers on $S^n$. In this lower dimensional dual field theory one can now reverse this process, and consider Fourier transforms
\ben
\cO_{y^i} (x^\mu) \equiv \int d^mk ~\cO_{k_i} (x^\mu)~e^{i\vk \cdot \vy}.
\een
which is of the form of (\ref{0-1}). Here we have adopted a notation which emphasizes that from the point of view of the IR field theory $y^i$ should be considered as a label at this stage. The discussion of RT surfaces then strongly suggest that the piece of the entanglement entropy which is evaluated by a RT surface in the IR geometry with a cutoff, which is anchored on a region of the internal space, relates to the subalgebra of these operators with $y^i$ restricted to the appropriate region.

While these considerations do not apply to purely product spaces which are not IR geometries of a higher dimensional $AdS$ space, the above discussion motivates us to probe properties of operators of the form (\ref{0-1}) in these cases as well. In this paper we ask the question: in what sense can we regard the $y^i$ as a label for a point in the internal space ? We do not expect them to be local in the internal space in the sense of characteristic singularities of e.g. two point functions for coincident points. In fact, the above discussion shows that we need to consider such product spaces with a (bulk) IR cutoff, corresponding to a UV cutoff in the dual field theory. Indeed, we will argue that this is essential to make sense of an expansion like (\ref{0-1}), since the different terms have dimension conformal dimension. However, we can ask if these operators obey clustering properties, e.g. whether the two point function of operators which are far separated in the internal space and separated by at least the cutoff in the base space decay. In this paper we address this question by computing two point functions of scalar operators like (\ref{0-1}). 

Throughout this paper, we will consider euclidean two point functions. We will work in units where the $AdS$ scale is unity. For flow geometries the scale of the lower dimensional $AdS$ space is related to the UV $AdS$ scale by a factor of order one.

We first consider the CFT which is dual to a product space $EAdS_{d+1-n} \times Y^n$ where $Y^n$ is a space with an isometry group $G$. This means that the R symmetry of the CFT is G. The scalar primary operators form a representation of this symmetry, characterized by some set of quantum numbers which we will continue to denote by $(l,\vm)$, and the dimensions of these operators are given by the standard AdS/CFT correspondence. The operators of interest are of the form (\ref{0-1}) where $Y_{l,\vm} (\theta_i)$ form a irreducible representation of $G$. As mentioned above, we need to properly define the operator sums like (\ref{0-1}) by introducing a scale so that all terms in the sum have the same dimension. We will do that by inserting an appropriate factor of the UV cutoff of the field theory, as would naturally follow from the bulk. We will show, in complete generality, that starting from the standard expression for the two point function in the CFT, the correlator of these operators can be expressed in terms of the Euclidean propagator of an auxilliary massive scalar field on (Euclidean time)~$\times Y^n$. The Euclidean time coordinate (which we will denote by $u$)  is the logarithm of the (invariant) distance between the two points in the CFT in units of the UV cutoff. The mass of this auxilliary scalar (which we will denote by $M$) is related to the conformal dimension of the lowest primary and is order one in $AdS$ units. The final result will be a function of the geodesic distance (Euclidean time)~$\times Y^n$. When $Y^n$ is non-compact (e.g. $R^n$), the correlator therefore falls off exponentially for geodesic distances large compared to $1/M$. This implies that there is clustering in the internal space. When $Y^n$ is compact with a size $R_n$ in AdS units, he same conclusion would hold if the separation in the internal space is much smaller than $R_n$ but much larger than  $1/M$, which is possible for $R_n \gg 1/M$. 

It should be emphasized that this is the result of a calculation in the CFT - the input from AdS/CFT is the sprectrum. AdS/CFT also provides a preferred normalization of the two point function of the primaries: we show that this normalizaion does not spoil the long distance behavior described above. We then perform a bulk calculation of the same correlator in the geodesic approximation, vindicating the above result. 

We then consider a bulk calculation in a geometry which flows from a $AdS_{d+1}$ in the UV to $AdS_{d+1-n} \times R^n$. Such a flow is governed by a scale $r_h$ which acts a cutoff of the IR geometry. For ease of calculation we consider a geometry with the metric
\ben
ds^2 = \left( \frac{dr^2}{r^2} + r^2 \sum_{i=1}^{d-n} (dx_i)^2 \right) + (r+r_h)^2\sum_{j=1}^n (dy_j)^2
\label{0-2}
\een
This is a carricature of a genuine solution in supergravity (e.g. RN black holes or magnetic branes).
However this is good enough to illustrate the main point. In particular, this enables us to treat arbitrary $d$ and $n$ in a unified fashion.  It turns out the scalar wave equation can be exactly solved in this metric.
However, since we are interested in the IR limit, we will calculate the correlator first by solving the wave equation for small energy-momentum in the field theory directions which become part of $AdS_{d+1-n}$ along the lines of \cite{liu} (see also \cite{other}). The main motivation for this detailed calculation is to determine the normalization of the IR correlator, which would be different from pure product space. Generically we indeed find that the IR correlator behaves as that in a product space geometry: the momenta along the $R^n$ now appear as quantum numbers for an internal space. The different normalization does not spoil the long distance behavior. 

The main lesson which we can draw from these calculations is the following. When the internal space is non-compact, there is approximate locality in the internal space at scales much larger than $1/M$, which is of order one in AdS units. When the internal space is compact the exponential fall-off of the correlator can set in only when there is a separation of the scale of the size of the compact directions compared to the AdS scale. In standard realizations of AdS/CFT in String/M theory this does not happen: the operators of the form (\ref{0-1}) are not local in any useful sense. However, in the cases where the product spaces appear as a IR geometry of a higher dimensional AdS space, there is such a separation of scale and locality in the internal space. For example in the geometry (\ref{0-2}) this is $r_h$. In extremal RN black holes, this relative scale is the black hole charge, while for magnetic branes this is the magnetic field. This makes sense since these flow geometries are dual of RG flows of the field theory, and the IR theory should be local for distance scales larger than the scale of the flow.

Finally, we comment on the role of color degrees of freedom in constructing operators of the form (\ref{0-1}) and then restricting them to some subregion of the internal space. The operators $\cO_{l,\vm}$ are of course gauge invariant operators.  However, in typical realizations of AdS/CFT these are single traces of products of $N \times N$ matrix valued operators. Because of trace relationships these are not independent at any finite $N$; thus this geomerization of the internal space can happen only in the $N \rightarrow \infty$. We will also argue that the count of the number of degrees of freedom which underlie holography \cite{susskind-witten} itself implies that a restriction to some region in this internal space must come from a restriction on the color degrees of freedom.

The question of (approximate) locality in an internal space which appears in the IR of a flow geometry $AdS_{d+1} \rightarrow AdS_{2} \times R^{d-1}$ has been considered earlier in \cite{iqbal,iqbal2} in the context of condensed matter applications of AdS/CFT. Our results are more general, involving arbitrary internal spaces and $AdS$ factors of arbitrary dimensionality. Furthermore the physical context as well as the lessons are quite different.

In section (\ref{product}) we calculate the exact two point correlators in product spaces from the CFT point of view and provide the result of a bulk geodesic calculation in the geodesic approximation . In section (\ref{flow}) we discuss correlators in the flow geometry (\ref{0-2}). This contains a detailed solution of the wave equation in the low (invariant) energy limit obtained by matched asymptotic expansions. The procedure is along the lines of \cite{liu}. In section (\ref{lesson}) we discuss the lesson drawn from these calculations and the implications for the question of entanglement of internal degrees of freedom. Section (\ref{color}) deals with the role of color degrees of freedom. In an appendix we comment on the appearence of locality in target space in other contexts.

\section{Correlators in a product space} 
\label{product}

In this section we study product spaces of the form $AdS_{d+1-n} \times Y^n$ in Euclidean signature. Here $Y^n$ is some n-dimensional internal space. The metric is given by
\ben
ds^2 = \frac{1}{z^2} \left[  d\vx^2  + dz^2 \right] + 
R_n^2 g_{ij}(y) dy^i dy^j
\label{1-1}
\een
where $y^i, i = 1 \cdots n$ are the coordinates of the internal space and $\vx =(x^0 \cdots x^{d+1-n})$ and $z$ are (Poincare patch) coordinates on $AdS_{d+1-n}$. The $AdS$ scale has been set to unity. $R_n$ is then the scale of the internal space in these units.

The internal space has some isometry group, which we will call R-symmetry.
The primary operators in the dual field theory living on the boundary of the $AdS_{d+1-n}$ at $z=0$ are labelled by the quantum numbers of the R symmetry which we will generically denote by $(l, \vm)$. For example, when the internal space is a $S^n$, these are angular momentum quantum numbers.
When the internal space is a $R^n$ the quantum numbers are the momenta $k_i$ along the $R^n$ directions. 
Consider a scalar operator $\cO_{l,\vm}(x^\mu)$ which is dual to a bulk scalar field with mass $m_0$. This has a conformal dimension $\Delta_\ell$ 
\ben
\Delta_\ell =\frac{d-n}{2} + \nu_{\ell}~~~~~~~~~~\nu_{\ell} = \sqrt{ \frac{(d-n)^2}{4} + m_0^2 + \ell^2}
\label{1-2}
\een
where $\ell^2$ denote the eigenvalues of the Laplacian on $Y^n$,
\ben
\nabla^2_{Y}Y_{\ell,\vm} (\vy) = \ell^2 Y_{\ell,\vm} (\vy)
\een
For example, when the internal space is $S^n$ we have $\ell^2 = l(l+n-1)/R_n^2$ and $Y_{\ell,\vm}$ are the spherical harmonics on $S^n$. When it is $R^n$ we have $\ell^2 = \vk^2/R_n^2$.
These equations follow from a KK reduction of the bulk wave equation on $Y^m$. Starting with the wave equation in $(n+m+2)$ dimensions,
\ben
 \left[ \nabla^2_{AdS} + \nabla^2_{Y}+ m_0^2 \right]\phi (z,\vx, \vy) = 0
 \label{1-3}
 \een
we need to make to decompose the field into modes $\phi_{l,\vm}(z,\vx)$
\ben
\phi(z,\vx,\vy) = \sum_{l,\vm} ~Y_{\ell,\vm} (\vy)~\phi_{l,\vm}(z,\vx)
\label{1-4}
\een
so that
\ben
\left[ \nabla^2_{AdS} + \ell^2+ m_0^2 \right]\phi_{l,\vm}(z,\vx) = 0
\label{1-5}
\een
Then $\phi_{l,\vm}(z,\vx)$ are dual to the operators $\cO_{l,\vm}(x^\mu)$.

In analogy with the bulk decomposition (\ref{1-4}), we may be tempted to define an operator like (\ref{0-1}). However the different terms in the sum have different conformal dimension: this means that the expression (\ref{0-1}) as it stands needs revision. To make sense of this kind of sum we will work in $AdS$ with a cutoff at some small value of $z = \epsilon$. Then the field $\phi_{l,\vm}(z,\vx)$ has the asymptotics $\phi_{l,\vm}(z,\vx) \rightarrow \epsilon^{\Delta_\ell} \phi_{l,\vm,0}(\vx)$. This  $\phi_{l,\vm,0}(\vx)$ is the source which couples to the operator $\cO_{l,\vm}(x^\mu)$. This suggests that we consider the {\em un-renormalized} operators
\ben
 \tilde{\cO}_{l,\vm}(x^\mu) =
\epsilon^{\Delta_\ell}\cO_{l,\vm}(x^\mu)
\label{1-6}
\een
We can now combine them into an operator which is labelled by a point on the internal space as well as a point on the base space,
\ben
{\tilde{\cO}}(\vx,\vy) = \sum_{\ell,\vm} {\tilde{\cO}_{\ell,\vm}}(\vx)Y_{\ell,\vm} (\vy)
\label{1-7}
\een
Since $\epsilon$ is a cutoff in the CFT, distances in the base space which are smaller than $\epsilon$ do not make sense.  

\subsection{Exact calculation}
\label{exact}

We now calculate the correlators of ${\tilde{\cO}}(\vx,\vy)$. The calculations in this section are those in the CFT with no direct reference to the bulk. However the CFT we are considering has a spectrum of conformal dimensions given by (\ref{1-2}) which appears in a hologarphic CFT as described above.

The correlators of ${\tilde{\cO}_{\ell,\vm}}(\vx)$ are given by
\ben
\langle \tcO_{\ell,\vm} (\vx)\tcO_{\ell^\prime,\vm^\prime} (\vx^\prime) \rangle = C(\nu_{\ell} ) \left( \frac{\epsilon}{|\vx - \vx^\prime|} \right)^{2 \Delta_{\ell}}~\delta_{\ell,\ell^\prime}\delta_{\vm,-\vm^\prime}
\label{3-1}
\een
We have included a normalization factor $C(\nu_\ell)$. In a CFT this does not have any significance. We will, however, retain this for two reasons. First, if this is the IR CFT which flows from some other CFT in the UV, this facor is significant. Secondly, when we are dealing with a CFT which is dual to a bulk AdS, there is a natural identification of the bulk field at the boundary with the dual operator, and that provides a definite $C(\nu_\ell)$ \cite{mathur}, whose effect will be considered at the end of this section.
\ben
C(\nu_{\ell}) = \frac{(2\nu_{\ell})}{\pi^{\frac{d-n}{2}}} \frac{\Gamma(\frac{d-n}{2}+\nu_{\ell})}{\Gamma(\nu_{\ell})} 
\label{3-2}
\een

The correlators of operators defined in (\ref{1-7}) then follow from (\ref{3-1}),
\ben
\langle \tcO (\vx,\vy) \tcO (\vx^\prime, \vy^\prime) \rangle = e^{-\frac{d-n}{2}u}\sum_{\ell,\vm} 
C(\nu_{\ell}) ~{\rm exp} [-\nu_{\ell} u ]~Y_{\ell,\vm} (\vy)Y_{\ell,\vm} (\vy^\prime)
\label{3-5}
\een
where we have defined the quantity
\ben
u \equiv 2 \log \left( \frac{|\vx - \vx^\prime|}{\epsilon} \right)
\label{3-6}
\een
Since $\epsilon$ is the UV cutoff, we can only consider $|\vx - \vx^\prime| > \epsilon$ so that $u > 0$. 

The key observation is that this correlator is closely related to the Euclidean correlator of a massive scalar field on $Y^n \times~$ (Euclidean time).
To see this, use the identity
\ben
{\rm exp} [-\nu_{\ell} u ]~Y_{\ell,\vm} (\vy)Y_{\ell,\vm}(\vy^\prime) =  -\frac{\partial}{\partial u} D_{\ell,m} (u,\vy,\vy^\prime)
\label{3-7}
\een
where $D_{\ell,m} (u,\vy,\vy^\prime)$ denotes the contribution of the mode $(\ell,\vm)$ to the euclidean propagator of a scalar of mass 
\ben
M = \sqrt{\frac{(d-n)^2}{4} + m_0^2}
\label{3-7a}
\een
on the space $Y^n$ with $u$ playing the role of euclidean time,
\ben
D_{\ell,m} (u,\vy,\vy^\prime) = 
\int_{-\infty}^\infty \frac{d \omega}{2\pi i} \frac{e^{-i\omega u}}{\omega^2 + \ell^2 + M^2} Y_{\ell,\vm} (\vy)Y_{\ell,\vm}(\vy^\prime)
\label{3-8}
\een
Note that this scalar is not the same scalar we started out with. The full propagator of this scalar on $Y^n \times ({\rm time})$ is given by
\ben
D (u,\vy,\vy^\prime) = \sum_{\ell,\vm} D_{\ell,m} (u,\vy,\vy^\prime)
\label{3-9}
\een
and the correlator of interest may be formally written as
\ben
\langle \cO (\vx,\vy) \cO (\vx^\prime, \vy^\prime) \rangle = e^{-\frac{d}{2}u}
C(-\partial_u)~ \partial_u D (u,\vy,\vy^\prime) 
\label{3-10}
\een
The explicit form of $D (u,\vy,\vy^\prime)$ of course depends on the details of the internal space. However, the result will be function of the geodesic distance $\gamma$ between the two points in $Y^m \times$~(time),
\ben
\gamma^2 = u^2 + R_n^2 [s(\vy,\vy^\prime)]^2
\een
where $s(\vy,\vy^\prime)$ is the geodesic distance on $Y^n$ with unit size. 

When the internal space is non-compact the correlator decays exponentially for $\gamma > \frac{1}{M}$,
\ben
D (u,\vy,\vy^\prime) \sim {\rm exp}[ - M\gamma ]
\label{3-11}
\een
Thus there is clustering. In particular, for a given value of $u$, 

When the internal space is compact, an exponential behavior will set in for a range
\ben
1/M \ll R_n s \ll R_n
\een

The differential operator $C(-\partial_u)$ does not change this long distance behavior. This is because the coefficient $C(\nu_l)$ is finite in the $l \rightarrow 0$ limit and has a power series expansion in $\ell$. In fact when $(d-n)$ is odd, this factor is polynomial in $\partial_u$. 

An explicit expression for the correlator for $AdS_{d+1-n} \times R^n$ is given in Appendix A.

\subsection{Geodesic approximation}
\label{geodesic}

We now perform a bulk calculation of the correlator in a geodesic approximatiopn for a product space (\ref{1-1}). The idea is to calculate the bulk 2 point function of the scalar field dual to the operator in question $\langle \phi(z,\vx,\vy) \phi (z, \vx^\prime, \vy^\prime \rangle $ on a constant $z$ slice and take the limit $z = \epsilon$ at the end. When $m_0$ is large (in AdS units) a saddle point approximation holds 
\ben
\langle\phi(z, \vx, \vy)\phi(z,\vx^\prime, \vy^\prime)\rangle \sim \sum_{\text {geodesics }} e^{-m L}
\label{4-1}
\een
where $L$ is the geodesic distance between these points.
If there are multiple geodesics the leading answer is given by the geodesic with the smallest invariant length. The calculation is self-consistent when $m_0L \gg 1$.

Since we have a product space 
\ben
L^2 = L_{AdS}^2 + R_n^2 [s(\vy,\vy^\prime)]^2
\label{4-2}
\een
Here $L_{AdS}$ is the geodesic length between two points $(z,\vx)$ and $(z,\vx^\prime)$ in $AdS_{d+1-n}$, and $s(\vy,\vy^\prime)$ is the geodesic length in $Y^n$.  $L_{AdS}$ is given by the well known expression \cite{dhoker-freedman}
\ben
L_{AdS} = \log \{ \frac{1}{2z^2} \left[ 2z^2 + (\Delta \vx)^2 + \Delta \vx \sqrt{4z^2 + (\Delta \vx)^2 } \right] \}
\label{4-3}
\een
In the limit $z = \epsilon \ll 1$ this reduces to 
\ben
L_{AdS} \approx u
\label{4-4}
\een
where $u$ is defined in (\ref{3-6}), so that the correlator becomes
\ben
\langle\phi(\epsilon, \vx, \vy)\phi(\epsilon,\vx^\prime, \vy^\prime)\rangle \sim {\rm exp} \left[ - m_0 \sqrt{u^2 + R_n^2 [s(\vy,\vy^\prime)]^2}\right]
\label{4-5}
\een
Since we are working in the limit of large $m_0$, $M \sim m_0$ ( $M$ is defined in (\ref{3-7a})) This therefore reproduces the long distance behavior of the exact correlator in the previous subsection (\ref{3-11}). 

\subsection{Global AdS}

In the above expressions, the $AdS$ part of the metric is chosen to be the Poincare metric. In any other coordinate system the CFT correlator would be an expression similar to (\ref{3-1}) with the quantity $|\vx - \vx^\prime|$ replaced by the appropriate invariant distance in the boundary field theory. For example when the $AdS$ is in global coordinates
\ben
ds^2 = \sec^2\rho [ dt^2 + d\rho^2 + \sin^2\rho d\Omega_{d-n-1}^2 ]
\label{3-12}
\een
the expression ({\ref{3-1}) is replaced by
\ben
\langle \tcO_{\ell,\vm} (t,\theta_i)\tcO_{\ell,-\vm} (t^\prime, \theta_i^{\prime}) \rangle \sim \left( \frac{\delta}{d(t,\theta_i;t^\prime,\theta_i^\prime)} \right)^{2 \Delta_{\ell}}
\label{3-13}
\een
where $\theta_i$ denote the angles on the $S^{d-n-1}$ in (\ref{3-11}) and 
\ben
[d(t,\theta_i;t^\prime,\theta_i^\prime)]^2 =  4 \sinh^2 \left(\frac{t-t^\prime}{2} \right) +[ v (\theta_i,\theta_i^\prime)]^2
\een
$v (\theta_i,\theta_i^\prime)$ denotes the geodesic distance on the sphere $S^{d-n-1}$ at the cutoff boundary $\rho = \frac{\pi}{2} - \delta$. In the expression for the quantity $u$ in (\ref{3-6}) the distance $|\vx - \vx^\prime|$ is replaced by 
$d(t,\theta_i;t^\prime,\theta_i^\prime)$.

\section{Correlators in Flow geometries}
\label{flow}

In this section we evaluate correlators in a flow geometry and examine the precise limit in which their IR behavior is captured by product spaces. In particular we will consider a geometry which interpolates between $AdS_{d+1}$ and $AdS_{d+1-n} \times R^n$,
\ben
d s^2=  \left(\frac{dr^2}{r^2 } + r^2 \sum_{i}^{d-n} (d x_{i})^2\right) + (r+r_h)^2 \sum_{i}^n (dy_i)^2
\label{metric}
\een
This metric is not supposed to be a supergravity solution, but a convenient background to work with. The correlator of an operator $\cO (\vx,\vy)$ which is dual to a bulk scalar with mass $m_0$, $\langle \cO (\vx,\vy)\cO (\vx^\prime,\vy^\prime) \rangle$, should behave as a $d$ dimensional CFT correlator in the UV. In the IR, $|\vx -\vx^\prime| \gg r_h$ one expects to behave as a correlator in a $(d-n)$ dimensional CFT which has an internal target space. More specifically the Fourier transform in the $\vy$ space,
\ben
G(\vk,\vx - \vx^\prime) = \int d^n y ~\langle \cO (\vx,\vy)\cO (\vx^\prime,\vy^\prime) \rangle e^{i\vk \cdot \vy}
\label{4-1}
\een
should scale like a correlator of an operator with dimension 
\ben
\Delta_k = \frac{d-n}{2} + \alpha_k~~~~~~~~~~\alpha_k = \sqrt{\frac{(d-n)^2}{4}+\frac{k^2}{r_h^2} + m_0^2}
\label{4-2}
\een
In the following we will remove the subscript $k$ in $\alpha_k$.
The form of this IR correlator should be like (\ref{3-1}), with the scale $r_h$ playing the role of the cutoff $\epsilon$ and a coefficient which is different from the purely product space answer. If this coefficient has a nice $\vk \rightarrow 0$ limit,  we will have an exponential fall off for $r_h |\vy - \vy^\prime| \gg 1/M$ where $M$ is given by (\ref{3-7a}).

We will calculate the fourier transform
\ben
G(\vk,\vomega) = \int d^{d-n}x \int d^n y ~\langle \cO (\vx,\vy)\cO (0,0) \rangle e^{i\vomega \cdot \vx + i\vk \cdot \vy}
\label{4-1}
\een
by solving the wave equation in the low energy-momentum regime $|\vomega| \ll r_h$. This calculation is along the lines of \cite{liu}. 

To leading order we will show that 
\begin{equation}
G(\omega,k) =K \big(2\nu r_h^{2\nu} \big) \frac{\G(-2\nu)}{\G(2\nu)} \ \frac{\G(\A - \B + \nu)}{\G(\A -\B - \nu)} \ \frac{\G(1+\A + \B + \nu)}{\G(1 + \A + \B - \nu)} \Bigg[ 1 + b(\A, \B, \nu) \left( \frac{\omega}{2r_h}\right)^{2\A} + \ldots \Bigg],
\label{5-1}
\end{equation}
where $K$ is positive constant coming from the normalization of the action and the coefficient $b(\A, \B, \nu)$ is
\begin{align}
b(\A, \B, \nu) = \frac{\G(-\A) \G(1-2\A) }{\G(\A) \G(1+2\A) } \Bigg\{ & \frac{\G(\A-\B-\nu)}{\G(-\A +\B+1-\nu)}  \frac{\G(\A+ \B +1 - \nu)}{\G(-\A-\B-\nu)}  \nonumber\\ 
& - ~\frac{\G(\A-\B+\nu)}{\G(-\A +\B+1+\nu)} \frac{\G(\A+ \B +1 + \nu)}{\G(-\A-\B+\nu)} \Bigg\},
\label{5-2}
\end{align}
with $\omega \equiv |\vec{\omega}|$, $k \equiv |\vec{k}|$, 
\begin{equation}\label{eqn:parameters}
\alpha={\sqrt{\frac{(d-n)^2}{4}+ m_0^2+\frac{k^2}{r_h^2}}},~~ \beta = \sqrt{ \frac{(n-1)^2}{4} + \frac{k^2}{r_h^2} } - \frac12 ,\text{ and } 
\nu = \sqrt{\frac{d^2}{4}+ m_0^2}.
\end{equation}
The leading non-analytic term in (\ref{5-1}), upon inverse Fourier transform in ${\vec{\omega}}$ for fixed $k$ , leads to a power law of the form (\ref{3-1}) with $\ell^2=k^2/r_h^2$. However the coefficient is different from (\ref{3-2}). These expressions clearly show that for $m_0 \neq 0$ these coefficients generically have a smooth limit as $k \rightarrow 0$.

We now proceed to a derivation of this result. 

\subsubsection{The Setup}

A free scalar field living on a background given by (\ref{metric}) obeys the Klein-Gordon equation  
\ben
{r}^{2}{\frac {{ d^{2}\Phi_{\vec{\omega},\vec{k}} (r)}}{{ dr}^{2}}} +{\frac {r \left(  \left( d-n+1 \right) r_h+ \left( 1+d \right) r\right)  }{r+r_h}}\frac { d\Phi_{\vec{\omega},\vec{k}} \left( r\right)}{dr} - \left( {\frac {{\vec{\omega}}^{2}}{{r}^{2}}}+{\frac {{\vec{k}}^{2}}{ \left( r+ r_h \right) ^{2}}}+{m_0}^{2} \right) \Phi_{\vec{\omega},\vec{k}} \left( r \right) =0
\label{KG}
\een
With
\ben
 \Phi(r,\vec{x},\vec{y})=\int \frac{d^{(d-n)} \omega~ d^n k}{(2\pi)^{d}} \Phi_{\vec{\omega},\vec{k}} (r)  e^{i(\vec{\omega}.\vec{x}+\vec{k}.\vec{y})}
\een
We can solve this wave equation exactly in terms of confluent Heun functions, and the two independent solutions are given as
\ben 
e^{-\frac{\omega}{r} } ~ r^{\pm \nu - \frac{d}{2}} ~ \left(1+ \frac{r_h}{r} \right)^{\beta + 1 - \frac{n}{2} } \text{HeunC} \left(\frac{2\omega}{r_h}, \mp 2\nu, 2\beta +1, 0, \frac{n(d-1)}{2} + \frac12, -\frac{r_h}{r}  \right)
\label{eqn:HeunSoln}
\een
%
%
The next step would be to impose regularity at $r=0$ which would enforce a certain linear combination of the two solutions. It turns out however, that obtaining an expansion of the confluent Heun functions as $r \rightarrow 0$ is rather difficult. As such, we will try to obtain a solution by matching the near horizon and asymptotic solutions instead.

\subsubsection{Matching solution}
A matching procedure for fields living on such flow geometries was developed in \cite{liu}. In this approach, one develops a careful perturbative expansion in powers of $\omega$. The central result is then that the full Green's function can then be written as \cite{liu}
\ben
G(\omega, k)=K \frac{b_{+}^{(0)}+\omega b_{+}^{(1)}+O\left(\omega^2\right)+\mathcal{G}_k(\omega)\left(b_{-}^{(0)}+\omega b_{-}^{(1)}+O\left(\omega^2\right)\right)}{a_{+}^{(0)}+\omega a_{+}^{(1)}+O\left(\omega^2\right)+\mathcal{G}_k(\omega)\left(a_{-}^{(0)}+\omega a_{-}^{(1)}+O\left(\omega^2\right)\right)},
\label{hong}
\een
where $a_{ \pm}^{(n)}$ and $b_{ \pm}^{(n)}$ are $k$-dependent functions, whose form only depend on the asymptotics of the full geometry, and $\mathcal{G}_k(\omega)$ is the IR region Green function. 

We will start by giving a short review of how in the case of a scalar field living in some flow geometry like the one described by \eqref{metric}, one can obtain an expression like \eqref{hong} for the Euclidean Green's function. The fundamental idea of this discussion will follow a similar line to the one given in \cite{liu} with a few adjustments and being adapted to our case of interest.  

We are interested in obtaining a solution for \eqref{KG} in the small $\omega$ regime that is regular at $r=0$. Let us denote this solution by $\phi$. Then one way to state the idea for the matching is roughly as follows: suppose $\phi$ admits a small $\omega$ series expansion, then our matching will be based on claiming that the operations of taking the limit $r/r_h \rightarrow 0$ and expanding in small $\omega$ commute . And so schematically we have 
\ben
\left(\text{Small $\omega$ expansion} \rightarrow \lim_{r/r_h \rightarrow 0} \right) \phi =  \left(\lim_{r/r_h \rightarrow 0} \rightarrow  \text{Small $\omega$ expansion}\right) \phi
\label{limit}
\een
More generally, the limit $r/r_h \rightarrow 0$ should be understood as the limit that, when applied to the Klein-Gordon equation, produces the near horizon limit. Throughout this section, in taking this limit we impose the condition that $\omega/r$ is kept finite.
The usefulness of the statement (\ref{limit}) is that one can try to solve \eqref{KG} in between these two operations After which, enforcing the above equality would be establishing the "matching" by making sure that the $\omega$ expansion matches on either side. 

A natural first approach to expand in small $\omega$ first would be to try to use perturbation theory to develop the solution as a power series in $\omega^2$. Such an approach however is bound to fail as $r \rightarrow 0$ since the $\omega^2$ term in \eqref{KG} blows up in this limit for any finite $\omega$. But not all is lost, perturbation theory is still a perfectly valid approach here as long as we stay away from $r=0$ and stop our solution at some cutoff $r=\epsilon$. The solution obtained this way then has the following form
\ben
\phi=C^+\left(\phi^+_0+\omega^2 \phi^+_1+ \ldots \right) +C^-\left(\phi^-_0+\omega^2 \phi^-_1+ \ldots \right) 
\label{expansion}
\een
where $\phi^\pm_0$ are the two linearly independent solutions one gets by plugging in $\omega=0$ in \eqref{KG}. The $\phi^\pm_i$ in \eqref{expansion} are then the $i$-th corrections to $\phi^\pm_0$ one gets from standard perturbation theory with the requirement that they do not have terms proportional to $\phi^\pm_0$. Now note that as long as the cutoff $\epsilon$ is not made equal to $0$, we can make it to be as small as we like. This implies that, at least formally, \eqref{expansion} can be made to represent the solution at any $r \neq 0$.  

To get the Euclidean Green's function however, we need to make our solution regular at $r=0$. Enforcing this condition then specifies what $C^\pm$ are in terms of the parameters of the problem, up to a freedom in choosing an overall normalization factor of the field. 

To determine $C^\pm$, we try taking the limit $r/r_h \rightarrow 0$ in \eqref{KG} first this time. This produces
 \ben
{r}^{2}{\frac {{ d^{2}\Phi_{\vec{\omega},\vec{k}} \left( r \right)}}{{ dr}^{2}}} + r  \left( d-n+1 \right)  \frac { d\Phi_{\vec{\omega},\vec{k}} \left( r\right)}{dr} - \left( {\frac {{\omega}^{2}}{{r}^{2}}}+{\frac {{k}^{2}}{  r_h ^{2}}}+{m_0}^{2} \right) \Phi_{\vec{\omega},\vec{k}} \left( r \right) =0
\label{productKG}
\een
which is just the equation describing a scalar field in a product space $AdS_{d-n+1} \times R^n$. The most general solution to \eqref{productKG} that is regular at $r=0$ is given by

\ben
 \Phi_{\vec{\omega},\vec{k}} \left( r\right)=B r^{\frac{n-d}{2}} K_\alpha\left(\frac{\omega}{r}\right) 
\een
where $B$ is some constant and $\alpha$ is given by \eqref{eqn:parameters}.

This solution of course has a standard expansion in $\omega$ given by
\ben
\Phi_{\vec{\omega},\vec{k}} \left( r\right)=B r^{\frac{n-d}{2}} \left[ 2^{\alpha-1} \Gamma(\alpha)\left(\frac{\omega}{r}\right)^{-\alpha}[1+\ldots]-2^{-\alpha-1} \frac{\Gamma(1-\alpha)}{\alpha}\left(\frac{\omega}{r}\right)^\alpha [1+\ldots]\right]
\label{asymptproduct}
\een
This should be then matched with the $r/r_h \rightarrow 0$ limit of \eqref{expansion}. For $\alpha$ non-integer, one only needs $\phi_0^\pm$ in \eqref{expansion} to completely determine $C^\pm$. Let us show this explicitly in our case.  The equation determining $\phi^\pm_0$ is simply given by

 \ben
{r}^{2}{\frac {{ d^{2}\phi^\pm_0 }}{{ dr}^{2}}} +{\frac {r \left(  \left( d-n+1 \right) r_h+ \left( 1+d \right) r\right)  }{r+r_h}}\frac { d\phi^\pm_0 }{dr} - \left( {\frac {{k}^{2}}{ \left( r+ r_h \right) ^{2}}}+{m_0}^{2} \right) \phi^\pm_0  =0
\label{zeroequation}
\een
This equation has an exact solution given by
\ben
 \phi^\pm_0=C^\pm r^{(\pm \alpha+\frac{n-d}{2})} (r+r_h)^{-(\beta+\frac{n}{2})}  F\left(\pm\alpha-\beta-\nu,\pm \alpha-\beta+\nu;1 \pm 2 \alpha;-\frac{r}{r_h}\right)
 \label{outersolution}
\een
where parameters $\alpha, \beta$ and $\nu$ are given by \eqref{eqn:parameters}. In the limit $r/r_h \rightarrow 0$, this expression simply becomes 
\ben
\phi^\pm_0=C^\pm r^{(\pm \alpha+\frac{n-d}{2})} r_h^{-(\beta+\frac{n}{2})}  
\label{outerr=0}
\een
Now if we had a situation with strictly $\omega=0$, then \eqref{outersolution} is really the whole solution and the regularity condition at $r=0$ simply sets $C^-=0$. For non-zero $\omega$, we match \eqref{outerr=0} with \eqref{asymptproduct} which then gives
\ben
C^\pm=B  2^{\pm\alpha-1}  r_h^{(\beta+\frac{n}{2})} \Gamma(\pm\alpha)\omega^{\mp\alpha}
\een
This then completes our matching with the solution now explicitly given by
\begin{multline}
 \phi=  2^{\alpha-1}   \Gamma(\alpha)\omega^{-\alpha}\left[ (r+r_h)^{-(\beta+\frac{n}{2})}r^{(\alpha+\frac{n-d}{2})} F\left(\alpha-\beta-\nu, \alpha-\beta+\nu;1+2 \alpha;-\frac{r}{r_h}\right) + \mathcal{O}\left(\omega^2\right)\right] \\ ~~~~~~ +  2^{-\alpha-1}   \Gamma(-\alpha)\omega^{\alpha} \left[ (r+r_h)^{-(\beta+\frac{n}{2})} r^{(-\alpha+\frac{n-d}{2})} F\left(-\alpha-\beta-\nu, -\alpha-\beta+\nu;1-2 \alpha;-\frac{r}{r_h}\right)+\mathcal{O}\left(\omega^2\right)\right]     
\end{multline}
where we have set $B r_h^{(\beta+\frac{n}{2})}=1$.

Now to derive an expression for the Green's function we need the asymptotic form of the solution as $r \rightarrow \infty$ . To this end, we make use of the relation 
\begin{multline}
 F(a, b ; c ; z)=\frac{\Gamma(b-a) \Gamma(c)}{\Gamma(b) \Gamma(c-a)}(-z)^{-a} F\left(a, a-c+1 ; a-b+1 ; \frac{1}{z}\right) \\ +\frac{\Gamma(a-b) \Gamma(c)}{\Gamma(a) \Gamma(c-b)}(-z)^{-b}F\left(b, b-c+1 ;-a+b+1 ; \frac{1}{z}\right)     
\end{multline}
Which is valid as long as the condition $ a-b\notin \mathbb{Z} \bigwedge z \notin(0,1)$ is met. Using this we can write the solution as 
\begin{multline*}
    \phi =2^{\alpha-1} \Gamma(\alpha)\omega^{-\alpha}  \Bigg[\bigg\{\frac{\Gamma(-2 \nu) \Gamma(1+2 \alpha)}{\Gamma(\alpha-\beta-\nu) \Gamma(1+\alpha+\beta-\nu)} (r+r_h)^{-\beta-\frac{n}{2}}r^{\alpha+\frac{n-d}{2}} \left(\frac{r}{r h}\right)^{-\alpha+\beta-\nu}   \\ ~~~~~~~~~~~~~~~~~~~~~~~~~~~~~~~~~~~~~~~~ F\Big(\alpha-\beta+\nu,-\alpha-\beta+\nu;1+2\nu;-\frac{r_h}{r}\Big) + \left(\nu \leftrightarrow -\nu \right) \bigg\} +\mathcal{O}(\omega^2)\Bigg] \\ ~~~~~~~~~~~ + 2^{-\alpha-1} \Gamma(-\alpha)\omega^{\alpha}  \Bigg[\bigg\{\frac{\Gamma(-2 \nu) \Gamma(1-2 \alpha)}{\Gamma(-\alpha-\beta-\nu) \Gamma(1-\alpha+\beta-\nu)} (r+r_h)^{-\beta-\frac{n}{2}}r^{-\alpha+\frac{n-d}{2}} \left(\frac{r}{r h}\right)^{\alpha+\beta-\nu}   \\ ~~~~~~ F\Big(-\alpha-\beta+\nu,\alpha-\beta+\nu;1+2\nu ;-\frac{r_h}{r}\Big) + \left(\nu \leftrightarrow -\nu \right) \bigg\} +\mathcal{O}(\omega^2)\Bigg]
\end{multline*}
This has an asymptotic expansion as $r \rightarrow \infty$ given by 
\begin{multline}
 \phi =\Bigg[\bigg\{ 2^{\alpha-1} \Gamma(\alpha)\omega^{-\alpha} \frac{\Gamma(1+2 \alpha) \Gamma(2 \nu)  r_h^{\alpha-\beta-\nu}}{\Gamma(\alpha-\beta+\nu) \Gamma(1+\alpha+\beta+\nu)}+ \left(\alpha \leftrightarrow -\alpha \right)\bigg\}\left(1+\mathcal{O}(\omega^2)\right)\Bigg] r^{-\frac{d}{2}+\nu} [1+\ldots] \\ +\Bigg[\bigg\{ 2^{\alpha-1} \Gamma(\alpha)\omega^{-\alpha} \frac{\Gamma(1+2 \alpha) \Gamma(-2 \nu)  r_h^{\alpha-\beta+\nu}}{\Gamma(\alpha-\beta-\nu) \Gamma(1+\alpha+\beta-\nu)}+ \left(\alpha \leftrightarrow -\alpha \right)\bigg\}\left(1+\mathcal{O}(\omega^2)\right)\Bigg] r^{-\frac{d}{2}-\nu} [1+\ldots]
\end{multline}
Then the Euclidean propagator of the boundary theory is simply given as the ratio of coefficient of the normalizable mode with that of the non-normalizable mode up to a normalization constant $K$. If we drop the higher order corrections in $\omega^2$ then the Euclidean propagator can be written as
\begin{equation}
G(\omega,k) = K \big(2\nu r_h^{2\nu} \big) \frac{\G(-2\nu)}{\G(2\nu)}\left[ \frac{   \frac{\G(\A) \G(1+2\A)}{\G(\A - \B - \nu) \G(1+ \A + \B - \nu)} + \left( \frac{\omega}{2 r_h} \right)^{2\A} \frac{\G(-\A) \G(1-2\A)}{\G(-\A - \B - \nu) \G(1-\A + \B - \nu)} }{ \frac{\G(\A) \G(1+2\A)}{\G(\A - \B + \nu) \G(1+ \A + \B + \nu)} + \left( \frac{\omega}{2 r_h} \right)^{2\A} \frac{\G(-\A) \G(1-2\A)}{\G(-\A - \B + \nu) \G(1-\A + \B + \nu)} } \right]
\label{maingreen}
\end{equation}
This result agrees with result given by the prescription in \cite{liu}. Since we have $\omega \ll r_h$, we can expand this expression as
\begin{equation}\label{eqn:Gsmallomega}
G(\omega,k) =K \big(2\nu r_h^{2\nu} \big) \frac{\G(-2\nu)}{\G(2\nu)} \ \frac{\G(\A - \B + \nu)}{\G(\A -\B - \nu)} \ \frac{\G(1+\A + \B + \nu)}{\G(1 + \A + \B - \nu)} \Bigg[ 1 + b(\A, \B, \nu) \left( \frac{\omega}{2r_h}\right)^{2\A} + \ldots \Bigg],
\end{equation}
where the coefficient $b(\A, \B, \nu)$ is
\begin{align}
b(\A, \B, \nu) = \frac{\G(-\A) \G(1-2\A) }{\G(\A) \G(1+2\A) } \Bigg\{ & \frac{\G(\A-\B-\nu)}{\G(-\A +\B+1-\nu)}  \frac{\G(\A+ \B +1 - \nu)}{\G(-\A-\B-\nu)}  \nonumber\\ 
& - ~\frac{\G(\A-\B+\nu)}{\G(-\A +\B+1+\nu)} \frac{\G(\A+ \B +1 + \nu)}{\G(-\A-\B+\nu)} \Bigg\} .
\end{align}
Here, the term containing $\mathcal{O}(\omega^{2\alpha})$ is the leading non-analytic piece, as the first term in the expansion (\ref{eqn:Gsmallomega}) contributes delta function in $\vec{x}$ when integrated over $\vec{\omega}$.

\section{Lessons for Entropies in Internal Spaces}
\label{lesson}

Our considerations were motivated by an attempt to understand entropies associated with subregions of internal spaces in product space geometries. In situations where such product spaces appear as IR geometries of higher dimensional asymptotically AdS spaces, we argued in \cite{dkkmrt} that such entropies are associated with subalgebras formed by sums and products of operators of the form (\ref{0-1}). This led us to investigate if the same continues to be the case for purely product spaces even if they do not have such UV completions. This would make sense if these operators have clustering properties in the internal directions. We found that these operators indeed cluster at a distance of the order of the $AdS$ scale. 
For compact internal spaces this implies that when the size of the internal space, $R_n$ is much larger than the $AdS$ scale we can consider these operators to be approximately local. 

Such a separation of scales is present when the product space has an UV completion as an asymptotically higher dimensional AdS space, as in the interpolating geometry (\ref{metric}). In more realistic geometries like extremal AdS-RN black holes the internal space is the horizon, and the size of the horizon is set by the charge and can be much larger than the AdS scale (the IR AdS scale is related to the higher dimensional AdS scale by a factor of order unity), and these operators cluster in a meaningful sense. Indeed this is what we expect from the corresponding RG flow in the dual field theory. If we consider the entropy of a subregion of the UV boundary theory which covers some part of the internal space and is smeared along the other spatial directions, the operators which are associated with this entropy are precisely of the form (\ref{0-1}). An appropriate part of this entropy can be then associated with the putative dual of the IR product space geometry.

For realizations of holography in String/M-theory, e.g. $AdS_5 \times S^5$ and $AdS_4 \times S^7$, there is no such separation of scales since the size of the $AdS$ and internal spaces are both set by the scale of the flux which supports these solutions. The operators like (\ref{0-1}) cannot be considered to be local in any useful sense. In these cases, it is possible that the non-locality is related to the volume law for small subregions \footnote{This possibility has been suggested by T. Takayanagi (private communication)}. Indeed such a relationship is known for a class of non-local theories \cite{takanonlocal}.

The relationship between the relative size of the internal space $R_n$ to the $AdS$ scale and the appearance of clustering behavior can also be probed using $RT$ surfaces. From this point of view, this clustering behavior can manifest itself as a phase transition in the form of the dependence of the area of the $RT$ surface on the size of the region it is anchored on. For example, as we argued in \cite{dkkmrt}, the area of an $RT$ surface associated with an infinite strip anchored on the internal space of a product space geometry of the form $AdS_{d+1-n} \times R^n$ (with $d-n>1$) can only depend on the width of the strip up to a maximum width $l_{max} \sim 1/(d-n-1)$ \footnote{For magnetic branes this is a reflection of a gap produced by the external magnetic field \cite{kraus,luca}}. Whether strips of widths larger than $l_{max}$ then exist or not now depends on whether the internal space size is large enough to incorporate them. Once again, we see that this clustering behavior can only be seen in cases where the size of the internal space $R_n$ is larger than the $AdS$ scale.

\section{Role of color degrees of freedom}
\label{color}

From the point of view of the large-N CFT, the way internal spaces arise appears to be quite different from the way the radial direction of the AdS arises. The latter is thought to come from a scale in the field theory, while the former is a geometrization of an internal symmetry of the theory. In fact, the construction of the operators ${\tilde{\cO}}(\vx,\vy)$ in (\ref{1-7}) does not appear to involve any of the color degrees of freedom. In a similar vein, the interpretation  of RT surfaces anchored on a region of the internal space in terms of entanglement of a subset of the color degrees of freedom \cite{shiba} appears to be very different from the interpretation of \cite{ku} in terms of operators like (\ref{1-7}). On the other hand one might expect that the color degrees of freedom do play a key role. For example, we certainly do not expect that construction of these operators make any sense when the number of colors is small. In this section we comment on this issue and argue that in fact the two proposals \cite{shiba} and \cite{ku} should be the same in spirit.

Let us first see why a large number of colors are essential in constructing operators like ${\tilde{\cO}}(\vx,\vy)$. The expansion which defines these operators involve a sum over the quantum numbers of the internal space and {\em assumes} that in the dual field theory operators with different quantum numbers are independent of each other. This is false at any finite $N$. To see this, consider the specific case of a product space of the form $AdS_{5} \times S^5$. In the dual field theory the corresponding R symmetry of the dual theory is $SO(6)$. This R symmetry acts linearly on six scalars $\Phi^I, I = 1 \cdots 6$ each of which is a $N \times N$ matrix. The primary operators $\cO_{l,\vm} (\vx)$ are traces of symmetric traceless products of $\Phi^I$, e.g.
\ben
\cO_{l,\vm} (\vx) \sim {\rm Tr} (\Phi^{(I } (\vx) \Phi^J (\vx)\Phi^{K)} (\vx)\cdots - {\rm{trace}})
\label{7-1}
\een
The number of $\Phi$'s which appear is $l$. However the trace of a product of more than $N$ matrices is not independent of the traces of products of smaller number of matrices - rather they are related by trace relations. In fact the correct combination of traces are given by Schur Polynomials which are in one-to-one correspondence with Young diagrams, and the maximum number of rows of a Young diagram is $N$. Thus the maximum value of $l$ which can appear in the sum (\ref{1-7}) is of order $N$. The emergent internal space $\vy$ is fundamentally "grainy" and can be sometimes thought of as a noncommutative space. This fact is well known in matrix models \cite{jevicki} and plays a key role in the physics of giant gravitons and related states \cite{giant} and is often called the "Stringy Exclusion Principle" \cite{exclusion}. A smooth internal space can possibly arise at $N = \infty$ when the trace relations can be ignored. This fact is also essential in the finiteness of target space entanglement \cite{target1}-\cite{mandal}.

The fact that entanglement of regions of the internal space associated with operators like ${\tilde{\cO}}(\vx,\vy)$ should involve entanglement of color degrees of freedom in fact follows from a count of the degrees of freedom required by the holographic principle following \cite{susskind-witten}. Consider e.g. $AdS_5 \times S^5$ with the metric
\ben
ds^2 = R^2 \left[ \frac{1}{z^2}[-dt^2 + d\vx^2] + d\Omega_5^2 \right]
\label{7-2}
\een
where $R = R_{AdS}=R_5$ is the common radius of $AdS_5$ and $S^5$. Consider a cutoff boundary at $z = z_0$. The total area $A$ of this $R^{3,1}\times S^5$ cutoff boundary is 
\ben
A = R^8 \frac{L^3}{z_0^3}
\label{7-3}
\een
where $L$ is the extent of each of the spatial directions along $\vx$.
The holographic principle then requires that the associated entropy
\ben
S = \frac{A}{G_{10}}
\een
where $G_{10}$ is the ten dimensional Newton's constant,
provides a count of the number of degrees of freedom of the field theory. Using the relationships 
\ben
G_{10}\sim g_s^2 l_s^8~~~~~~~~(R/l_s)^4 \sim g_s N
\een
(where $l_s$ is the string length) this leads to the prediction for the number of degrees of freedom $N_{dof}$
\ben
N_{dof} \sim N^2 \frac{L^3}{z_0^3}
\label{7-4}
\een
which is the correct count once $z_0$ is identified with the UV (position space) cutoff of the dual field theory. This is simply the fact that there are $N^2$ degrees of freedom at every lattice site of the field theory on $R^{3,1}$.

In the above relations, the factor $R^8$ came from the eight spatial directions of the ten dimensional metric composed of $\vx$ and the angles in $S^5$. Now suppose we want to obtain the number of degrees of freedom associated with a fraction $f$ of the $S^5$, keeping the entire $R^3$. This count will be
\ben
N_{dof}^\prime \sim  f N^2 \frac{L^3}{z_0^3}
\label{7-5}
\een
This relationship is naturally interpreted as the contribution of $fN^2$ degrees of freedom at each lattice site. 
This is in the same spirit of \cite{shiba}. Note that we are dealing with {\em gauge invariant} degrees of freedom. It will be interesting to find a precise gauge invariant meaning of such a decomposition of the color degrees of freedom.

There are other ways of thinking of entanglement of color degrees of freedom.
Target space entanglement is a gauge invariant way of formulating entanglement in color space \cite{target1}. The idea here is to consider a single hermitian combination of the matrix fields of the field theory and define a matrix valued projector which projects onto the subspace defined by restricting the eigenvalue of this hermitian matrix to be in a certain range. The projected operators and their products and sums define a subalgebra with an associated reduced density matrix. These ideas can be applied to e.g. subregions of a $S^5$ which emerges out of the six scalar fields $\Phi^I$  in the $3+1$ dimensional $N=4$ field theory. Consider for example a projector associated with the constraint that the eigenvalues of $\Phi^1$ are greater than some real number $a$. This may be thought about a restriction to a polar cap of a $S^5$ and the resulting target space entanglement may be thought of an entanglement in internal space. One situation where this kind of matrix entanglement can be made precise appears in \cite{hartnoll2} and \cite{frenkel2}. They consider e.g. a fuzzy sphere state of a three matrix problem. In such a state a projector associated with $\phi^1 > a$ leads to a target space entanglement which is proportial to the length of the boundary of the cap in units of the coupling constant of the emergent noncommutative $U(1)$ theory on the fuzzy 2-sphere. In \cite{frenkel2} it was found that this result is in fact quite general and applies to general regions in other non-commutative backgrounds generated in quantum mechanics of arbitrary number of matrices. It is as yet unclear if this is related to a RT surface. It will be interesting if this has a gravity dual description when embedded in e.g. BFSS model. Other notions of matrix entanglement can be defined in e.g. partially deconfined states \cite{hanada}.

\section{Acknowledgements} S.R.D would like to thank A. Frenkel, J. Gauntlett, S. Hartnoll, H. Lin and T. Takayanagi for discussions. The work of H.B., S.R.D and  M.H.R are partially supported by a National Science Foundation grant NSF-PHY/211673 and by the Jack and Linda Gill Chair Professorship. S.R.D. would like to thank Tata Institute of Fundamental Research, Higgs Center for Theoretical Physics, King's College (London), Yukawa Institute for Theoretical Physics and Isaac Newton Institute for Mathematical Sciences, U. of Cambridge for hospitality during various stages of this work.
GM, KKN and SPT acknowledge support from Government of India, Department of Atomic Energy, under Project Identification No. RTI 4002 and from the Quantum
Space-Time Endowment of the Infosys Science Foundation.

\appendix

\section{Correlator for $AdS_{d+1-n}\times R^n$}

In this section we give an explicit expression for the correlator in $AdS_{d+1-n}\times R^n$ following the subsection (\ref{exact}).

When the internal space is $R^m$ the eigenfunctions of the Laplacian on the internal space are plane waves
\ben
Y_\vk = \frac{1}{(2\pi)^{n/2}} e^{i R_n \vk \cdot \vy}
\label{3-12}
\een
Then the function $D(u,\vy,\vy^\prime)$ is given by
\bea
D^{(R^n)} (u,\vy,\vy^\prime) & = &\int \frac{d^n k}{(2\pi)^n} \int \frac{d\omega}{2\pi}
\frac{1}{\omega^2 + \vk^2+ M^2}e^{-i\omega u +i R_n \vk \cdot (\vy - \vy^\prime)} \nonumber \\
& = & \left(\frac{M}{\sqrt{u^2 + R_n^2 (\Delta \vy)^2}}\right)^{\frac{n-1}{2}}
K_{\frac{n-1}{2}}(M\sqrt{u^2 + R_n^2 (\Delta \vy)^2})
\label{3-13}
\eea
where we have denoted $\Delta \vy \equiv \vy - \vy^\prime$ and $K$ denotes the modified Bessel function. The quantity
$\sqrt{u^2 + R_n^2(\Delta \vy)^2}$ is  the invariant distance on $({\rm (time)} \times R^n$. The large argument limit of the modified Bessel function is of course an exponential, leading to
\ben
D^{(R^n)} (u,\vy,\vy^\prime) \sim \frac{M^{(n-2)/2}}{(u^2 + R_n^2(\Delta \vy)^2)^{n/2}}{\rm exp} (- M \sqrt{u^2 + (\Delta \vy)^2})
\label{fall-off}
\een

\section{Other examples of ``target space locality''}\label{world-sheet}

It is interesting to speculate whether the notion of target space locality considered in the text appears in other contexts. In this Appendix, we will give one more example of this phenomenon.

Consider a 2D CFT with $d$ massless scalar fields (i.e. a CFT with a target space ${\mathbb R}^d$). We have the well-known primary fields
\begin{align}
  & O_k(z,\bz)= e^{ik.X(z,\bz)}, \quad
  \lan O_k(z,\bz) O_k^*(w,\bw) \ran = \left(\fr{|z-w|}{\epsilon}\right)^{-k^2}
  = \exp[-k^2 g(z,w)] \nonumber\\
  & g(z,w)= \log\left(\fr{|z-w|}{\epsilon}\right), \quad k \in {\mathbb R}^d
  \label{ws-2pt}
\end{align}
Let us define a ``local'' operator in target space:
\begin{align}
 & O(y, z,\bz)=\int \fr{d^d k}{(2\pi)^d} e^{-i k.y}O_k(z,\bz) = \delta^d(y-X(z,\bz)), \nonumber\\
 & \lan O(y,z,\bz) O(y',w,\bw) \ran = g(z,w)^{-d/2} \exp[- \fr{|y-y'|^2}{g(z,w)}]
\label{ts-2pt}
\end{align}
For fixed $(z,\bz), \, (w,\bw)$, the operators $O(y, z,\bz)$ satisfy ``cluster decomposition property'', in target space ${\mathbb R}^d$, i.e. the connected correlator $\lan O(y, z,\bz) O(y',w,\bz)\ran$ falls off for $y$ and $y'$ sufficiently far apart (for fixed $(z,\bz), \, (w,\bw)$). The behaviour \eq{ts-2pt} is reminscent of \eq{fall-off} or \eq{3-11}, except that at large distances the fall-off in \eq{ts-2pt} is more like in case of diffusion or a non-relativistic field theory.

The example above is closely related to the world sheet conformal theory of a string in ${\mathbb R}^d$. It is worthwhile remembering that in the context of a string the ``target space'' \underbar{is} the space time (with $d=26$ for bosonic string).\footnote{In the static gauge, the target space becomes the $d-2$ dimensional space orthogonal to the world sheet.} A more familiar manifestation of {\it target space locality} in the string context is the fact that vertex operator correlators, when integrated over the worldsheet, can be represented by an effective action (see, e.g. \cite{fradkin-tseytlin}) which is local in target space.\footnote{In the examples considered in the main body of the paper, there is no counterpart of the notion of integrating over the ``world sheet'', though.}

Similar remarks can also be made for D-branes moving in a target space. The case of D1 branes may be regarded as the closest to that of the fundamental string because of S-duality. However, it is true of more general D branes too; after all, the coupling of D branes to target space is given by a local (DBI) action. E.g. the target space of D3 branes is $M= R^6 (\times$  time, described by
a DBI action local in $M$. Of course, locality in $R^6$ does not imply locality in $S^5$ which is part of the $R^6$; here, input from the Maldacena limit plays a crucial role.

\end{document}